\documentclass[%
 reprint,
 amsmath,amssymb,
 aps,
prb,
floatfix,
]{revtex4-1}

\usepackage{graphicx}% Include figure files
\usepackage{dcolumn}% Align table columns on decimal point
\usepackage{bm}% bold math
\usepackage{multirow}
\usepackage{array}

\begin{document}

\preprint{APS/123-QED}

\title{Lossless and loss-induced topological transitions of isofrequency surfaces of a~composite magnetic-semiconductor medium}

\author{Volodymyr I. Fesenko$^{1,2}$}
\author{Vladimir R. Tuz$^{1,2}$}
\email{tvr@jlu.edu.cn; tvr@rian.kharkov.ua}
\affiliation{$^1$International Center of Future Science, State Key Laboratory of Integrated Optoelectronics, College of Electronic Science and Engineering, Jilin University, 2699 Qianjin St., Changchun 130012, China}
\affiliation{$^2$Institute of Radio Astronomy of National Academy of Sciences of Ukraine, 4, Mystetstv St., Kharkiv 61002, Ukraine} 

\date{\today}

\begin{abstract}
Topological transitions of isofrequency surfaces of a composite magnetic-semiconductor structure influenced by an external static magnetic field are studied in the long-wavelength approximation. For the lossless case, the topological transitions of isofrequency surfaces from a closed ellipsoid to open Type I and Type II hyperboloids as well as a bi-hyperboloid are demonstrated. Conditions for critical points where the topological transitions occur are found out. It is revealed that actual material losses in the constituents of the composite medium strongly influence the dispersion behaviours for the extraordinary waves, which manifest themselves in the loss-induced topological transitions of isofrequency surfaces. It is shown that the loss-induced topological transitions from a Type I hyperboloid to a bi-hyperboloid appear in the frequency band where the real part of a particular principal component of the anisotropic constitutive parameter (permittivity or permeability tensor) is a near-zero value while its imaginary part is a non-zero value.
%\begin{description}
%\end{description}
\end{abstract}

 %PACS, the Physics and Astronomy Classification Scheme.
%78.20.Ek Optical activity
%78.20.Fm Birefringence
%78.67.Pt Multilayers; superlattices; photonic structures; metamaterials

%\keywords{Suggested keywords} %Use show keys class option if keyword display desired

\maketitle
%\tableofcontents
\section{\label{intr}Introduction}
During last decade hyperbolic metamaterials are a subject of intense study due to their specific dispersion features that are unattainable in conventional media.\cite{Ferrari_2015} For today they are realized for operating in the microwave, terahertz, and optical ranges, and such unusual effects as the negative refraction,\cite{Cortes:2012} strong enhancement of spontaneous emission,\cite{Poddubny_PhysRevB_2013} broadband infinite density of states,\cite{Jacob_APL_2012} subwavelength imaging,\cite{Liu_Science_2007} focusing,\cite{Smith_APL_2004} and signal routing,\cite{Kapitanova_NatComm_2014} have been experimentally demonstrated.

Hyperbolic dispersion appears in \textit{extremely} anisotropic media (also known as \textit{indefinite} media\cite{Smith_PhysRevLett_2003}). In such media, at least one constitutive parameter necessarily is a tensor quantity $\hat{\eta} = \bigl[\eta_{xx}, 0, 0; 0, \eta_{yy}, 0;~0, 0, \eta_{zz} \bigr]$ (here $\eta$ is either permeability $\mu$ or permittivity $\varepsilon$), while one of the principal components of the tensor $\hat{\eta}$ has the opposite sign with respect to the rest principal components. A medium is considered to be hyperbolic uniaxial crystal or hyperbolic biaxial crystal when either the condition  $\eta_{xx}<0<\eta_{yy}=\eta_{zz}$ or the condition $\eta_{xx}<0<\eta_{yy}<\eta_{zz}$ between the tensor's diagonal components holds. In the indefinite media, the topological transitions of isofrequency surfaces appear when the real part of a particular component of permittivity or permeability tensor changes sign from positive to negative or vice versa. 

Typically, hyperbolic dispersion is sought in non-magnetic metamaterials characterized by an indefinite permittivity tensor and scalar permeability ($\mu_{xx} = \mu_{yy} = \mu_{zz} = 1$). Since the negative value of permittivity is an ordinary property of metals in the plasmonic conditions, hyperbolic metamaterials are usually constructed by incorporating metallic components into the structure designs. In such designs the metamaterials are made either in the form of superlattices which combine metallic and dielectric layers\cite{Schilling_PhysRevE_2006, Zhukovsky_OptExpress_2013, Poddubny2013} or lattices of metallic wires embedded into dielectric hosts.\cite{Simovski_AdvMat_2012, Sun_OE_2013} Regardless of design, the effective medium limit on the sizes of structural components  is usually implied for all configurations, therefore the overall structure is considered under the long-wavelength approximation.\cite{Kidwai_PhysRevA_2012} Additionally, hyperbolic metamaterials can be implemented utilizing the indefinite permeability tensor. This feature appears in a narrow frequency band for metamaterials composed of metallic split-ring resonators.\cite{Pendry_MTT_1999}   

Nevertheless, the hyperbolic dispersion exists also in natural media featuring gyroelectric (e.g., plasma or semiconductor) or gyromagnetic (e.g., ferrite) conditions, when the medium is influenced by an external static magnetic field. Under saturated magnetization such media become extremely anisotropic in a specific frequency band due to manifestation of the plasma or ferromagnetic resonance. It leads to appearance of hyperbolic isofrequency behaviours for particular waves (as is conventional in optics of anisotropic media,\cite{Fisher_PhysRevLett_1969, Kuznetsov_OptComm_2017, Lokk2017, Chern:17} the waves whose propagation is influenced by the external static magnetic field are related to the \textit{extraordinary} waves, otherwise they are \textit{ordinary} waves). In the general case of a biaxial gyroelectric or gyromagnetic medium there are topological transitions of isofrequency surfaces for the extraordinary waves from a closed ellipsoid ($0<\eta_{xx}<\eta_{yy}<\eta_{zz}$) to open Type~I ($\eta_{xx}<0<\eta_{yy}<\eta_{zz}$) or Type~II ($\eta_{xx}<\eta_{yy}<0<\eta_{zz}$) asymmetric hyperboloid.\cite{Sun_OE_2013, Ballantine2014} Regarding ordinary waves, their dispersion characteristics are typical with isofrequency surfaces being in the form of a closed ellipsoid. Thus, of greater interest are characteristics of the extraordinary waves since the topological transitions in their isofrequency surfaces from a closed form to open ones strongly modifies the propagation conditions of the waves, that can be utilized for manipulations by the matter-field interactions.\cite{Krishnamoorthy_Science_2012}

The topological transitions of isofrequency surfaces are usually studied separately for either gyroelectric \cite{Kuznetsov_OptComm_2017} or gyromagnetic media.\cite{Lokk2017, Chern:17} Nevertheless, it is demonstrated\cite{Kaganov_PhysUsp_1997,  Wu_JPhysCondensMatter_2007, Tarkhanyan_PSSb_2008, Shramkova_PIERM_2009, Tarkhanyan_JMMM_2010, Tuz_Springer_2016, Tuz_JMMM_2016, Fesenko_OptLett_2016, Tuz_JApplPhys_2017, Tuz_Superlattice_2017} that combining together semiconductor and magnetic materials into a unified gyroelectromagnetic structure (superlattice) whose permittivity and permeability are simultaneously tensor quantities gives unprecedented flexibly for management by dispersion features of both surface and bulk waves. In particular, in such a superlattice the topological transitions from a closed ellipsoid to open Type~I and Type~II hyperboloids as well as a bi-hyperboloid have been observed for the extraordinary waves.\cite{Tuz_OptLett_2017} The latter is a recently found form of topology of isofrequency surfaces which appears in the result of simultaneous tuning of both geometrical parameters of the superlattice and external static magnetic field influence. 

Typically, in order to simplify simulations and explanation of obtained results, the actual losses in materials forming a metamaterial are totally ignored assuming tensors $\hat{\varepsilon}$ and $\hat{\mu}$ are composed of only real values. In fact, such approximation is correct when considering conventional isotropic and anisotropic media ($\varepsilon_{ii} \ge 1$, $\mu_{ii} \ge 1$, $\varepsilon_{ij} = \mu_{ij} = 0$, $i \ne j$, $i,j = x,y,z$), in which presence of intrinsic losses leads only to the electromagnetic wave dumping during the wave propagation without any changing in the topological behaviours of dispersion.\cite{landau_1960_8, felsen1994radiation}

However, recently it has been reported\cite{Feng_PhysRevLett_2012} that presence of intrinsic losses in the $\varepsilon$-near-zero medium can play a positive role in the wave propagation. For instance, it can enhance transmission and collimate the field inside the medium. Moreover, it was experimentally observed \cite{Jiang_PhysRevB_2015, Yu_JAP_2016} that in the $\mu$-near-zero metamaterial constructed from two-dimensional transmission lines with lumped elements (resistors), a topological transition of isofrequency surfaces from a closed ellipsoid to an open Type~I hyperboloid appears. This topological transition was found when varying the imaginary part of permeability while both the real part of permeability and frequency are fixed. It  has been called \textit{loss-induced} topological transition, and it is qualitatively different from those appeared from changing the sign of the real part of permittivity or permeability. Several wave effects, namely collimation and field enhancement during the wave propagation were observed experimentally in the transmission-line-based metamaterial \cite{Jiang_PhysRevB_2015} supporting loss-induced transitions. The loss-induced transitions were also found in the two-dimensional transmission line metamaterials with an arbitrary positive real part of permeability\cite{Yu_JAP_2016} and in a one-dimensional periodic structure composed of graphene sheets deposited on dielectric layers.\cite{Guo_AppISci_2018}

In the present paper our main goal is to reveal characteristics of the topological transitions in isofrequency surfaces existing for the waves propagating through a gyroelectromagnetic structure. The existence of such transitions is expected since in the frequency bands near the plasma or ferromagnetic resonance the gyroelectromagnetic structure behaves as an $\varepsilon$-near-zero or $\mu$-near-zero medium. In these frequency bands, the intrinsic losses in constitutive materials are usually significant, and they can lead to the appearance of loss-induced topological transitions which are of our particular interest.  

The rest of the paper is organized as follows: In Section~\ref{Theory} we formulate and solve the problem related to the bulk waves propagating through a biaxial gyroelectromagnetic medium. Then in Section~\ref{Lossless}, we discuss the topological transitions of isofrequency surfaces in the idealized lossless biaxial gyroelectromagnetic medium and reveal the specific conditions under which they occur. In Section~\ref{Loss-induced} we extend the theory to account for intrinsic losses of actual materials forming the structure, and show the characteristics of loss-induced topological transitions appearing under certain geometrical and material parameters of the superlattice. Finally, in Section~\ref{Concl} we summarize the paper. The expressions for components of tensors characterizing ferrite and semiconductor materials are given in Appendix for reference.

\section{Dispersion relation for bulk waves}
\label{Theory}

\begin{figure*}[!ht]
\centering
\includegraphics[width=1.0\linewidth]{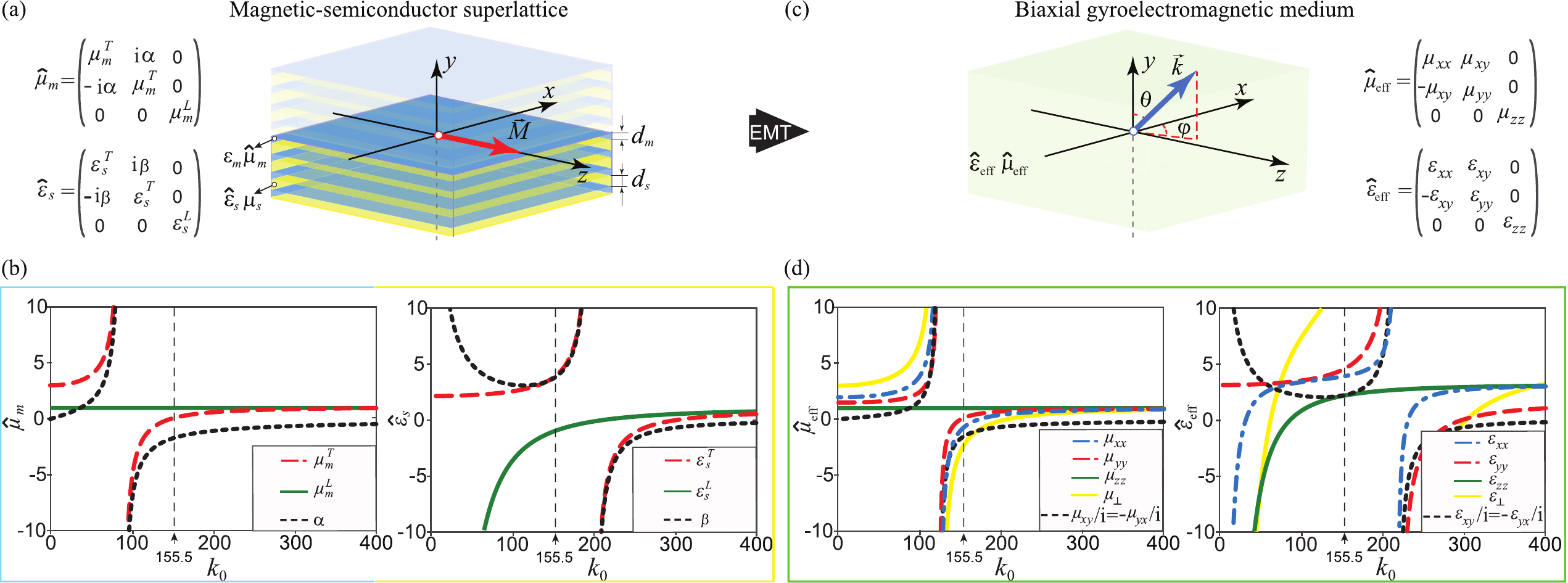}
\caption{The problem sketch related to: (a) a magnetic-semiconductor superlattice influenced by an external static magnetic field $\vec{M}$; (c) resulting homogenized medium (biaxial gyroelectromagnetic medium). Dispersion curves (real parts) of the tensor components of: (b) permeability $\hat \mu_m$ and permittivity $\hat \varepsilon_s$ of magnetic and semiconductor layers, respectively; (d) relative effective permeability $\hat \mu_{\textrm{eff}}$ and relative effective permittivity $\hat \varepsilon_{\textrm{eff}}$ of the homogenized medium with $\delta_m=\delta_s=0.5$. For the magnetic layers, under saturation magnetization of 2930~G, parameters are: $\omega_0/2\pi=4.2$~GHz, $\omega_m/2\pi=8.2$~GHz, $b=0$, $\varepsilon_m=5.5$. For the semiconductor layers, parameters are: $\omega_p/2\pi=10.5$~GHz, $\omega_c/2\pi=9.5$~GHz, $\nu/2\pi=0$~GHz, $\varepsilon_l=1.0$, $\mu_s=1.0$.}
\label{fig:fig1}
\end{figure*}

We study topological transitions of isofrequency surfaces related to a periodic (along the $y$-axis) multilayered structure (superlattice) of magnetic (ferrite) and semiconductor layers with thicknesses $d_m$ and $d_s$, respectively, that are infinitely extended along the transverse directions (Fig.~\ref{fig:fig1}(a)). The period of structure is $d = d_m + d_s$. All layers of the superlattice are magnetized uniformly by an external static magnetic field $\vec M$ which is directed along the $z$-axis transversely to the structure periodicity. Under the influence of such a magnetization, individual magnetic and semiconductor layers of the superlattice are characterized by the combination of tensor quantities of constitutive parameters $\hat \mu_m$, $\varepsilon_m$ and $\mu_s$, $\hat \varepsilon_s$. For clarity, in explicit view, the tensors $\hat \mu_m$ and $\hat \varepsilon_s$ and dispersion characteristics of their components are given in Fig.~\ref{fig:fig1}(b) (see expressions for components of tensors $\hat \mu_m$ and $\hat \varepsilon_s$ in Appendix).

We suppose that all characteristic dimensions of the superlattice, namely its period as well as the thicknesses of the constitutive layers satisfy the long-wavelength limit, that is $d_m\ll \lambda$, $d_s \ll \lambda$, $d \ll \lambda$, where $\lambda$ is the wavelength inside the medium. In such a way, the standard homogenization procedure from the effective medium theory (EMT) \cite{Agranovich_SolidStateCommun_1991} is applied to derive averaged expressions for effective parameters of the superlattice. As a result, the superlattice is equivalently substituted by infinite bigyrotropic space (Fig.~\ref{fig:fig1}(c)), which is characterized by relative effective permeability $\hat\mu_{\textrm{eff}}$ and relative effective permittivity $\hat\varepsilon_{\textrm{eff}}$ represented as the second-rank tensors (all details on the homogenization procedure are omitted here and can be found in Ref.~\onlinecite{Fesenko_book_2018}). After performing the homogenization procedure all principal components of tensors of permittivity as well as permeability appear to be different, namely $\varepsilon_{xx} \neq \varepsilon_{yy} \neq \varepsilon_{zz}$ and $\mu_{xx} \neq \mu_{yy} \neq \mu_{zz}$, thus the homogenized medium is a biaxial bigyrotropic ($\varepsilon_{xy} = -\varepsilon_{yx}\neq0$ and $\mu_{xy}=-\mu_{yx}\neq0$) crystal. In this crystal one of the optical axes is directed along the structure periodicity (the $y$-axis), while the second one coincides with the direction of the external magnetic field (the $z$-axis). It should be noted, that the anisotropy of the structure is strongly dependent on both the ratio between thicknesses of magnetic and semiconductor layers and the magnetic field strength. The corresponding dispersion characteristics of the components of relative effective permeability $\hat\mu_{\textrm{eff}}$ and relative effective permittivity $\hat\varepsilon_{\textrm{eff}}$ are given in Fig.~\ref{fig:fig1}(d) for reference.

In the following discussion we consider a plane uniform electromagnetic wave with angular frequency $\omega$ and wavevector $\vec{k}$ which propagates in a biaxial gyroelectromagnetic medium along an arbitrary direction as shown  in Fig.~\ref{fig:fig1}(c). The electric ($\vec E$)  and magnetic ($\vec H$) field vectors can be expressed as
\begin{equation}
\vec E (\vec H) = \vec E_0 (\vec H_0) \exp\left[\mathrm{i} \left(k_x x + k_y y + k_z z-\omega t\right)\right],
\label{eq:space_time}
\end{equation}
where $k_x=k \sin\theta \cos\varphi$, $k_y=k \sin\theta \sin\varphi$, and $k_z=k \cos\theta$ are Cartesian coordinates of the wavevector $\vec{k}$. 

From the Maxwell's equations the dispersion relation between $\omega=k_0c$ and $\vec{k}=\lbrace k_x,k_y,k_z \rbrace$ describing propagation of electromagnetic waves through an unbounded gyroelectromagnetic medium can be written as follow (the corresponding treatment is omitted here and can be found in Refs.~\onlinecite{landau_1960_8, Lokk2017}) 
\begin{equation}
\begin{split}
&(\varepsilon_{zz}\mu_{zz})^{-1}\biggl[k_x^4\varepsilon_{xx}\mu_{xx}+k_y^4\varepsilon_{yy}\mu_{yy}+k_z^4\varepsilon_{zz}\mu_{zz}+k_x^2k_y^2\times\\
&(\varepsilon_{xx}\mu_{yy}+\varepsilon_{yy}\mu_{xx})+k_x^2k_z^2(\varepsilon_{xx}\mu_{zz}+\varepsilon_{zz}\mu_{xx})+k_y^2k_z^2\times\\
&(\varepsilon_{yy}\mu_{zz}+\varepsilon_{zz}\mu_{yy})- 
k_0^2\biggl( k_x^2(\varepsilon_{xx}\varepsilon_{zz}\mu_{\bot}+\mu_{xx}\mu_{zz}\varepsilon_{\bot})+\\
&k_y^2(\varepsilon_{yy}\varepsilon_{zz}\mu_{\bot}+\mu_{yy}\mu_{zz}\varepsilon_{\bot})+k_z^2\varepsilon_{zz}\mu_{zz}(\varepsilon_{xx}\mu_{yy}+\varepsilon_{yy}\mu_{xx}\\
&-2\varepsilon_{xy}\mu_{xy}) \biggr) \biggl]+k_0^4\varepsilon_{\bot}\mu_{\bot}=0,
\end{split}
\label{eq:disp_eq}
\end{equation}
where $\varepsilon_{\bot} = \varepsilon_{xx}\varepsilon_{yy} + \varepsilon_{xy}^2$ and $\mu_{\bot} = \mu_{xx}\mu_{yy} + \mu_{xy}^2$ are two generalized transverse effective constitutive parameters. For clarity, the dispersion curves of parameters $\varepsilon_{\bot}$ and $\mu_{\bot}$ are shown in Fig.~\ref{fig:fig1}(d) by solid yellow lines.

At a constant frequency bi-quadratic dispersion equation (\ref{eq:disp_eq}) is the Fresnel's equation of wave normals\cite{landau_1960_8} and describes a fourth-order (\textit{quartic}) surface in the $k_x$, $k_y$, and $k_z$ coordinates which is known as an isofrequency surface (or, alternatively, the Fresnel wave surface or surface of wave vectors). In normalized terms of $\kappa=k/k_0$ it can be rewritten as\cite{Tuz_OptLett_2017}
\begin{equation}
A \kappa^4 + B\kappa^2+C=0,
\label{eq:disp_eq_k}
\end{equation}
whose solution is straightforward: $\kappa^2= (B \pm \sqrt{B^2-4AC})/2A$. 
Here, $A=(\varepsilon_{zz}\mu_{zz})^{-1} \left(\overline{\varepsilon} \sin^2\theta + \varepsilon_{zz} \cos^2\theta \right) \left(\overline{\mu} \sin^2\theta + \mu_{zz} \cos^2\theta\right)$, $B=-[(\varepsilon_{xx}\mu_{yy} + \mu_{xx}\varepsilon_{yy} - 2\varepsilon_{xy}\mu_{xy})\cos^2\theta +(\varepsilon_{zz}\mu_{zz})^{-1}(\varepsilon_{\bot}\overline{\mu}\mu_{zz}+\mu_{\bot}\overline{\varepsilon}\varepsilon_{zz})\sin^2\theta]$, $C=\varepsilon_{\bot}\mu_{\bot}$, $\overline{\varepsilon} = \varepsilon_{xx}\cos^2\varphi + \varepsilon_{yy}\sin^2\varphi$, and $\overline{\mu} = \mu_{xx}\cos^2\varphi + \mu_{yy}\sin^2\varphi$.

We should note, that equations (\ref{eq:disp_eq}) and (\ref{eq:disp_eq_k}) are different representations of the dispersion equation for bulk waves propagating in an unbounded biaxial gyroelectromagnetic medium. The solution of this dispersion equation gives us characteristics of the electromagnetic field inside the structure for the known set of geometrical and material parameters of the superlattice, so the \textit{direct} problem is formulated and solved here. Recently an elegant solution of the corresponding \textit{inverse} problem has been reported.\cite{Mulkey_OL_2018}    

When a lossless medium is supposed, then the isofrequency surfaces are related to the purely real roots of Eq.~(\ref{eq:disp_eq_k}). In the general case, two isofrequency surfaces exist at a particular frequency, it means that Eq.~(\ref{eq:disp_eq_k}) yields two real roots $\kappa_1$ and $\kappa_2$. In accordance with accepted notation,\cite{felsen1994radiation} one of such roots (with upper sign `$+$') is related to the \textit{ordinary} waves, while another one (with lower sign `$-$') is related to the \textit{extraordinary} waves. When one of the roots $\kappa_1$ or $\kappa_2$ is a purely imaginary quantity (which corresponds to propagation of an evanescent wave) only one isofrequency surface exists at the given frequency.  

Furthermore, in a lossy medium, the components of effective constitutive tensors $\hat{\varepsilon}_{\textrm{eff}}$ and $\hat{\mu}_{\textrm{eff}}$ are complex quantities, so, $\varepsilon_{ij}=\varepsilon_{ij}'+\mathrm{i}\varepsilon_{ij}''$  and $\mu_{ij}=\mu_{ij}'+\mathrm{i}\mu_{ij}''$, where  $i,j=x,y,z$. In this case, all four roots of Eq.~(\ref{eq:disp_eq_k}) are complex quantities (that is, $\kappa_i=\kappa'_i+\mathrm{i}\kappa''_i$, $i=1,2,3,4$) which represent propagation of complex waves. Their amplitudes either decay or grow exponentially during the wave propagation. Complex waves whose amplitude decays exponentially ($\kappa_i''> 0$) are related to the \textit{proper} waves,\cite{Ishimaru_book_1991, Pengchao_NP_2018} otherwise ($\kappa_i'' < 0$) they  are related to the \textit{improper} waves, which are nonphysical solutions for a passive medium.  %Further in our study, when a lossy medium is supposed, we are only interested  in the solutions of Eq.~(\ref{fig:fig3}) with a positive imaginary part, which describe propagation of the proper waves.

\section{Lossless topological transitions}
\label{Lossless}

Our objective here is to study the topological behaviours of isofrequency surfaces for the waves propagating through an idealized lossless gyroelectromagnetic medium. Nevertheless, excluding the losses, we set all other parameters of the given superlattice to values typical for actual materials. In particular, in this paper we follow the results of Ref.~\onlinecite{Wu_JPhysCondensMatter_2007}, and study the superlattice made of barium-cobalt and doped-silicon layers. This choice is due to the fact that the characteristic resonant frequencies for these materials being under the influence of an external magnetic field are, although different, sufficiently closely spaced within the same frequency band (see Fig.~\ref{fig:fig1}(b)). All geometrical and material parameters of the superlattice as well as the external magnetic field strength are summarized in the caption of Fig.~\ref{fig:fig1}. For all our subsequent calculations the frequency parameter is fixed at $k_0 = 155.5$~m$^{-1}$ ($f\approx 7.4$~GHz).

In the  superlattice under study, the externally applied static magnetic field can essentially change the dispersion characteristics of the medium since it simultaneously influences both permeability and permittivity of magnetic and semiconductor layers, respectively. The dispersion characteristics depend also on the direction of structure periodicity and ratio between filling factors of the magnetic and semiconductor constituents of the superlattice (we introduce corresponding filling factors as dimensionless parameters written in the form $\delta_m = d_m/d$, $\delta_s = d_s/d$, and $\delta_m + \delta_s = 1$). Nevertheless, it is revealed\cite{Tuz_OptLett_2017} that the topological transitions of isofrequency surfaces appear due to the changes in the principal values of tensors of relative effective permeability $\hat\mu_{\textrm{eff}}$ and relative effective permittivity $\hat\varepsilon_{\textrm{eff}}$. At a constant external magnetic field strength, these parameters remain to be functions of the superlattice filling factors only. 

\begin{figure}[!t]
\centering
\includegraphics[width=1.0\linewidth]{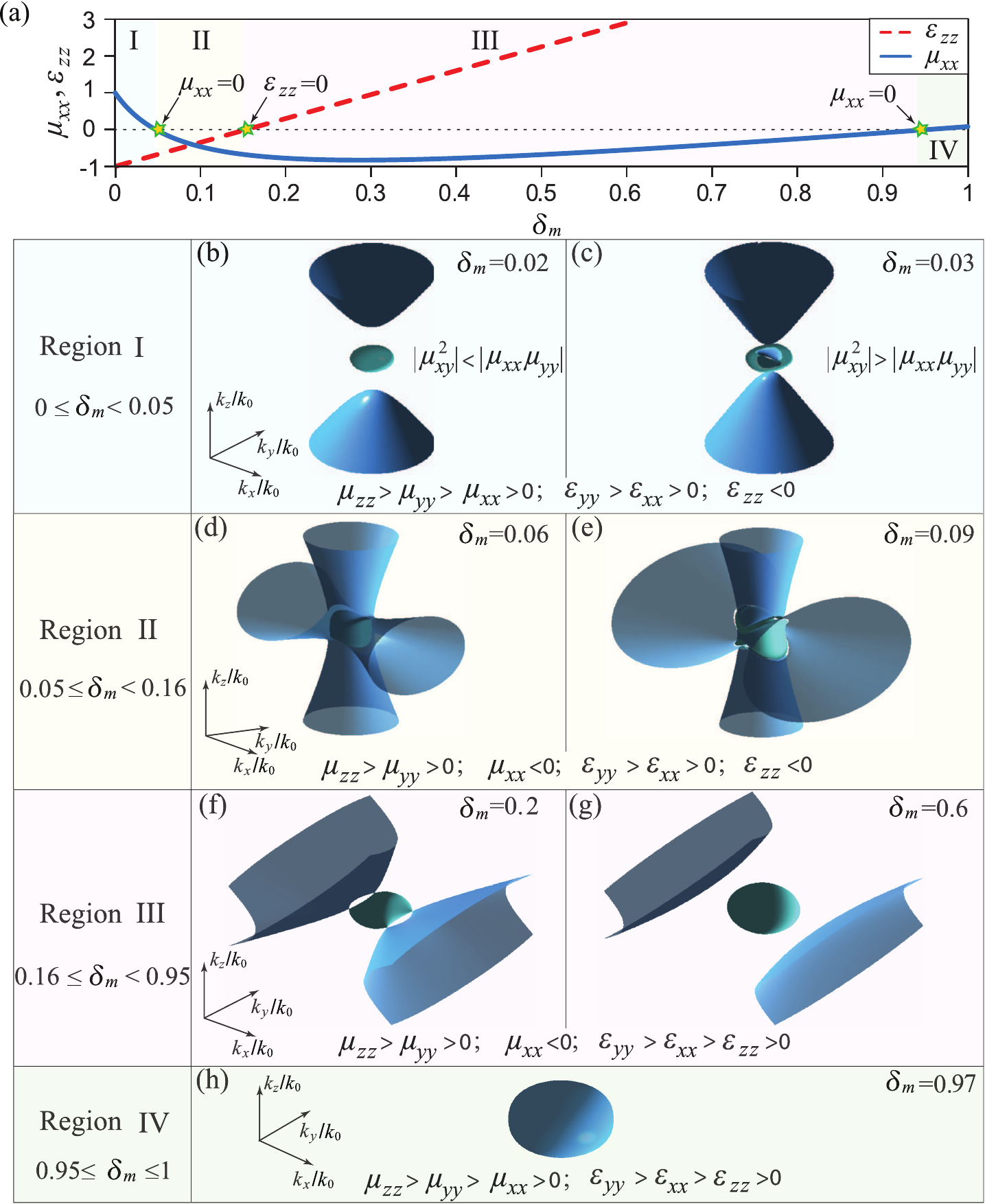}
\caption{(a) The change in the values of principal components $\mu_{xx}$ and $\varepsilon_{zz}$ as the parameter $\delta_m$ varies at a fixed frequency. (b)-(h) The forms of isofrequency surfaces related to waves propagating through a lossless superlattice for different values of the parameter $\delta_m$. Green and blue surfaces correspond to behaviours of ordinary and extraordinary waves, respectively.}
\label{fig:fig2}
\end{figure}

\begin{figure}[!t]
\centering
\includegraphics[width=1.0\linewidth]{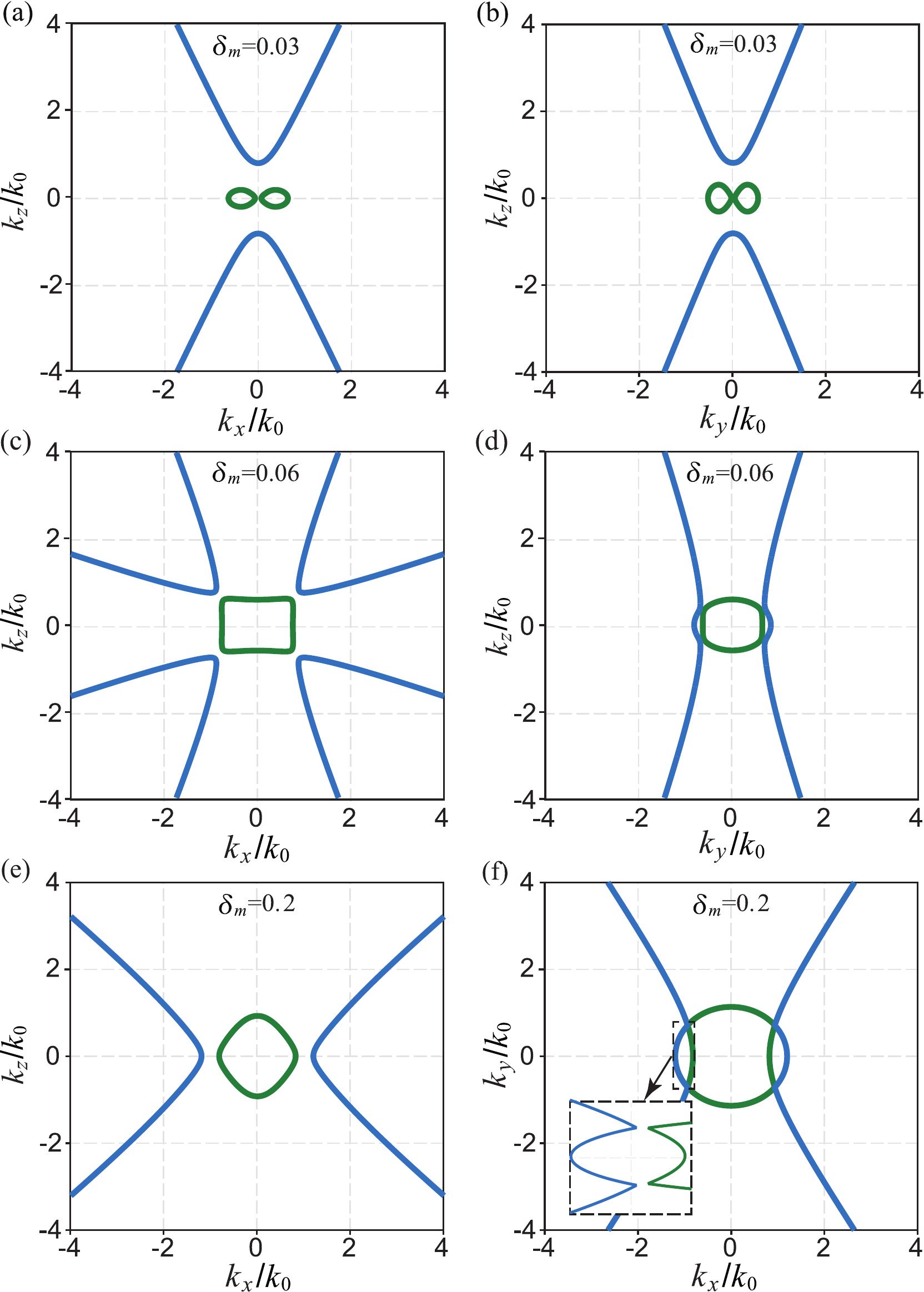}
\caption{The cross-section views of the isofrequency surfaces for different values of the parameter $\delta_m$. In the inset of Panel (f) disappearing of self-intersection points is demonstrated. Green and blue curves correspond to behaviours of ordinary and extraordinary waves, respectively.}
\label{fig:fig3}
\end{figure}

For the given parameters of the superlattice, the following relations between the values of principal components of tensors $\hat\mu_{\textrm{eff}}$ and $\hat\varepsilon_{\textrm{eff}}$ hold: $\mu_{zz}>\mu_{yy}>\mu_{xx}$ and $\varepsilon_{yy}>\varepsilon_{xx}>\varepsilon_{zz}$. The principal values $\varepsilon_{yy}$, $\varepsilon_{xx}$, and $\mu_{yy}$ are always positive quantities, and $\mu_{zz} = 1$ in the whole range of variation of the parameter $\delta_m$ ($\delta_m \in [0.0, 1.0]$). Moreover, several representative regions can be distinguished where different combinations of positive and negative values of principal components $\varepsilon_{zz}$ and $\mu_{xx}$ appear as the parameter $\delta_m$ varies (corresponding regions are denoted by Roman numerals `I-IV' and painted in different colors in Fig.~\ref{fig:fig2}(a)). They are: (Region~I) $\delta_m \in [0.0, 0.05)$, where $\varepsilon_{zz}<0$ and $\mu_{xx}>0$; (Region~II) $\delta_m \in [0.05, 0.16)$, where $\varepsilon_{zz}<0$ and $\mu_{xx}<0$; (Region~III) $\delta_m \in [0.16, 0.95)$, where $\varepsilon_{zz}>0$ and $\mu_{xx}<0$; (Region~IV) $\delta_m \in [0.95, 1.0]$, where $\varepsilon_{zz}>0$ and $\mu_{xx}>0$. 

These different combinations of values inevitably manifest themselves in the distinct forms of isofrequency surfaces which arise as a result of some topological transitions. These transitions appear when either value $\varepsilon_{zz}$ or $\mu_{xx}$ reaches zero as the parameter $\delta_m$ varies, that separates topologically distinct sets of solutions of Eq.~(\ref{eq:disp_eq_k}) (we depict the transition points in Fig.~\ref{fig:fig2}(a) by stars). Therefore, we plot the corresponding set of isofrequency surfaces in Figs.~\ref{fig:fig2}(b)-\ref{fig:fig2}(h) for Regions I-IV, respectively, where in the $k$-space each component is normalized on $k_0$. In the plots green and blue surfaces correspond to behaviours of ordinary and extraordinary waves, respectively.

In Region~I all principal components of $\hat\mu_{\textrm{eff}}$ are positive quantities ($|\mu_{xy}^2|<|\mu_{xx}\mu_{yy}|$), whereas in $\hat\varepsilon_{\textrm{eff}}$ the component $\varepsilon_{zz}$ is a negative one. The isofrequency surfaces arise in the forms of an ellipsoid for the ordinary waves and a two-fold Type~I uniaxial-hyperboloid for the extraordinary waves (Fig.~\ref{fig:fig2}(b)), as is typical for the hyperbolic metamaterials.\cite{Poddubny2013} Nevertheless, there is a distinctive feature consisting in the fact that the hyperboloid is slightly compressed along the $z$-axis, since in this direction there is an additional anisotropy axis as a result of the influence of an external magnetic field. Such a composite feature of isofrequency surfaces (known as the \textit{mixed-type} dispersion\cite{Chang_OE_2014}) is conditioned by the hybrid character of Eq.~(\ref{eq:disp_eq_k}) and appears when hyperbolicity in the permittivity tensor and gyrotropy in the permeability tensor exist simultaneously.\cite{Chang_OE_2014, Chern:17} Moreover, from Figs.~\ref{fig:fig3}(a) and \ref{fig:fig3}(b) one can conclude that the rotation symmetry of the isofrequency surfaces about the $z$-axis is broken that is typical for a biaxial crystal.\cite{Ballantine2014}

As the parameter $\delta_m$ increases, the ellipsoid-like surface gradually deforms by decreasing the thickness at the ellipsoid center performing transition to a toroid. Once the condition $|\mu_{xy}^2|=|\mu_{xx}\mu_{yy}|$ is achieved the thickness of ellipsoid at the center reduces to zero as presented in Figs.~\ref{fig:fig3}(a) and \ref{fig:fig3}(b). Subsequently, in the part of Region I, when the condition $|\mu_{xy}^2|>|\mu_{xx}\mu_{yy}|$ holds, the isofrequency surface of the ordinary waves transits to a toroid-like form as shown in Fig.~\ref{fig:fig2}(c). The similar dispersion behaviours have been recently reported for a hyperbolic-gyromagnetic metamaterial.\cite{Chern:17}

When both principal components $\varepsilon_{zz}$ and $\mu_{xx}$ become negative quantities (Region II), there is a topological transition of the isofrequency surface of the extraordinary waves to a bi-hyperbolic topology, that is presented by a combination of two one-fold Type II hyperboloids whose revolution axis are orthogonal (Figs.~\ref{fig:fig2}(d), \ref{fig:fig2}(e) and \ref{fig:fig3}(c)). Such an unusual bi-hyperbolic topology is a result of simultaneous effects of both the structure periodicity and influence of an external magnetic field.\cite{Tuz_OptLett_2017} The isofrequency surface of the ordinary waves transits to an ellipsoid-like form, since the condition $|\mu_{xy}^2|<|\mu_{xx}\mu_{yy}|$ is satisfied. This ellipsoid lies inside of such a complicated bi-hyperbolic form. 

\begin{figure*}[!t]
\centering
\includegraphics[width=1.0\linewidth]{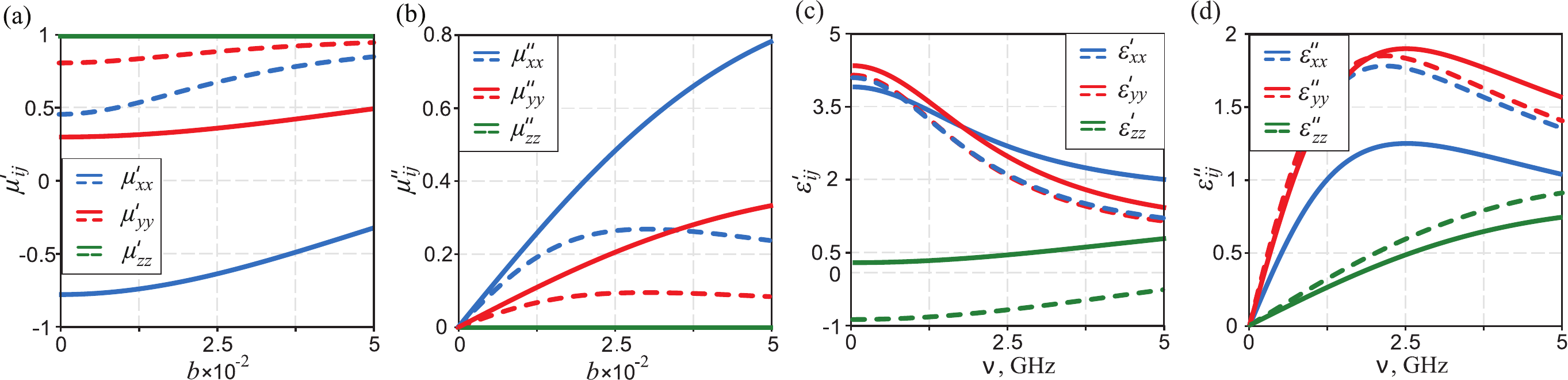}
\caption{Real and imaginary parts of components of tensors (a), (b) $\hat\mu_{\textrm{eff}}$ and (c), (d) $\hat\varepsilon_{\textrm{eff}}$ as functions of the dimensionless parameter $b$ and parameter $\nu$ given in GHz for two fixed values of the parameter $\delta_m$. Dashed and solid lines correspond to $\delta_m=0.02$ and $\delta_m=0.2$, respectively.  Other parameters are the same as in Fig.~\ref{fig:fig1}.}
\label{fig:fig4}
\end{figure*}

For the third set of parameters (Region III), among all principal values of relative effective tensors only $\mu_{xx}$ is a negative quantity. There is a topological transition of the isofrequency surface of the extraordinary waves to a Type I biaxial-hyperboloid oriented along the $x$-axis (Fig.~\ref{fig:fig2}(f)). As the parameter $\delta_m$ further increases this form gradually transits to a Type I uniaxial-hyperboloid as shown in Fig.~\ref{fig:fig2}(g). In these conditions the isofrequency surfaces are strongly compressed along the $z$-axis due to the influence of an external static magnetic field. An another effect of this field is a complete removing the degeneracy points (also known as \textit{self-intersection} points \cite{landau_1960_8}) in the isofrequency surfaces (Fig.~\ref{fig:fig3}(f)). In turn, it leads to disappearance of the conical refraction existed in anisotropic crystals,\cite{Kuznetsov_OptComm_2017,Ballantine2014} since the isofrequency surfaces split apart in these self-intersection points under an action of the external magnetic field.\cite{Khatkevich_1989,Belsky_2002} 

Finally, in Region IV ($\delta_m \gg \delta_s$) all principal components of $\hat{\varepsilon}_{\textrm{eff}}$ and $\hat{\mu}_{\textrm{eff}}$ are positive quantities, the isofrequency surface is in the form of a single ellipsoid as shown in Fig.~\ref{fig:fig2}(h). In this region $\varepsilon_{xy} \to 0$,  $\varepsilon_{xx}\approx\varepsilon_{yy}\approx\varepsilon_{zz} \to \varepsilon_m$, and the dispersion behaviours of a gyromagnetic (ferrite) bulk medium are dominant, for which the topological transitions are well described (see, for instance, Ref.~\onlinecite{Lokk2017} and references therein).

\section{Loss-induced topological transitions}
\label{Loss-induced}

The topological transitions and forms of isofrequency surfaces presented so far have been limited to the lossless case, but if the operating frequency is positioned closely to the characteristic frequency of either ferromagnetic or plasma resonance, the intrinsic losses in constitutive materials of the superlattice are significant and thus cannot be ignored. In lossy media, components of the wavevector $\vec{k}$ are complex quantities ($k_i = k_i' + \mathrm{i}k_i''$, $i = x, y, z$, where $k_i'$ represents the phase change, and $k_j''$ corresponds to the decay rate), therefore with accounting for losses some modifications in forms of isofrequency surfaces become possible.\cite{Ballantine2014, Yu_JAP_2016, Guo_AppISci_2018} These modifications are of our further study, and especially we are interested in the conditions of occurrence of the loss-induced topological transitions of the isofrequency surfaces. To this end, we again consider previously distinguished regions of variation of the parameter $\delta_m$, namely we have chosen the set of parameters: $\delta_m=0.02$, $\delta_m=0.06$, and $\delta_m=0.2$ which are representative values for Region I (Fig.~\ref{fig:fig2}(b)), Region II (Fig.~\ref{fig:fig2}(d)), and Region III (Fig.~\ref{fig:fig2}(f)), respectively. Since in Region IV there are no peculiarities, it is excluded from our further consideration. 

In our model parameters $b$ and $\nu$ are responsible for losses in magnetic and semiconductor layers, respectively (see Appendix). The parameter $b$ is dimensionless, whereas the parameter $\nu$ has the dimension of frequency. For clarity of discussion, the variance of both the real and imaginary parts of principal components of tensors $\hat{\mu}_{\textrm{eff}}$ and $\hat{\varepsilon}_{\textrm{eff}}$ conditioned by increasing $b$ and $\nu$ is shown in Fig.~\ref{fig:fig4} for two chosen values of the parameter $\delta_m$. 

\begin{figure}[!t]
\centering
\includegraphics[width=0.95\linewidth]{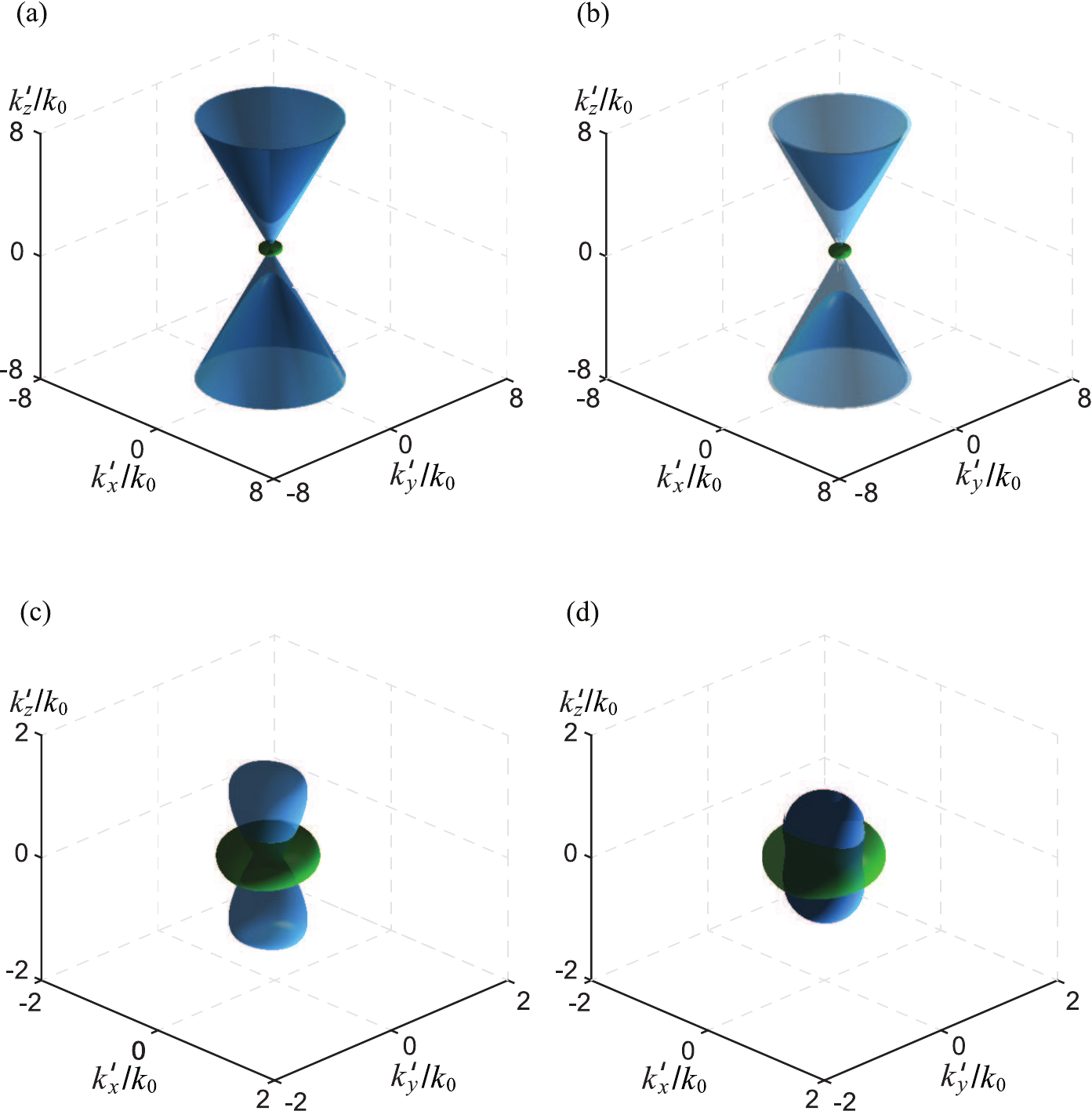}
\caption{The forms of isofrequency surfaces related to waves propagating through a superlattice characterized by different intrinsic losses which are typical for actual magnetic and semiconductor materials. For these materials corresponding conditions $-1 <\varepsilon_{zz}'<0$ and $0<\mu_{xx}'<1$ hold, and $\delta_m=0.02$. Parameters describing losses are: (a) $b = 1\times 10^{-4}$, $\nu = 1\times10^{-2}$~GHz; (b) $b = 5\times 10^{-2}$, $\nu = 1\times10^{-2}$~GHz; (c) $b = 1\times 10^{-4}$, $\nu = 2$~GHz; (d) $b = 5\times 10^{-2}$, $\nu = 2$~GHz.  Green and blue surfaces correspond to behaviours of ordinary and extraordinary waves, respectively.}
\label{fig:fig5}
\end{figure}

When $\delta_m=0.02$ (Region I) there is a combination  $0 < \mu_{xx}' < 1$ and $-1 < \varepsilon_{zz}' < 0$ for real parts of principal components of tensors $\hat\mu_{\textrm{eff}}$ and $\hat\varepsilon_{\textrm{eff}}$. Since $\mu_{xx}'$ is a small positive quantity, the superlattice behaves as a $\mu$-near-zero medium for which isofrequency surfaces have the forms of an ellipsoid and a Type I hyperboloid for ordinary and extraordinary waves, respectively (Fig.~\ref{fig:fig5}(a)). The revolution axis of the hyperboloid is oriented along the $z$-axis.

As soon as the discernible losses are introduced to the system ($\nu \neq 0$ and $b \neq 0$, that is $\mu_{ij}''\ne 0$ and $\varepsilon_{ij}''\ne 0$) the roots of Eq.~(\ref{eq:disp_eq_k}) becomes to be complex quantities for both ordinary and extraordinary waves. Their amplitude decays exponentially, so they belong to the proper waves. Since there is some attenuation in the system, the isofrequency surface of the extraordinary waves no longer goes to infinity, and the Type I hyperboloid becomes to be closed (Fig.~\ref{fig:fig5}(b)). The hyperbolic isofrequency surface of the extraordinary waves bends in the opposite direction at the finite value of $\vec{k}$ intersecting the ellipsoid-like isofrequency surface of the ordinary waves. In fact, the gradually increasing the level of losses leads to decreasing size of the closed hyperboloid-like area and finally to its disappearing. We should note, that such loss-induced topological transitions of the isofrequency surface of the extraordinary waves are well known for the hyperbolic metamaterials (see, for instance, Ref.~\onlinecite{Ballantine2014}).

\begin{figure}[!t]
\centering
\includegraphics[width=0.95\linewidth]{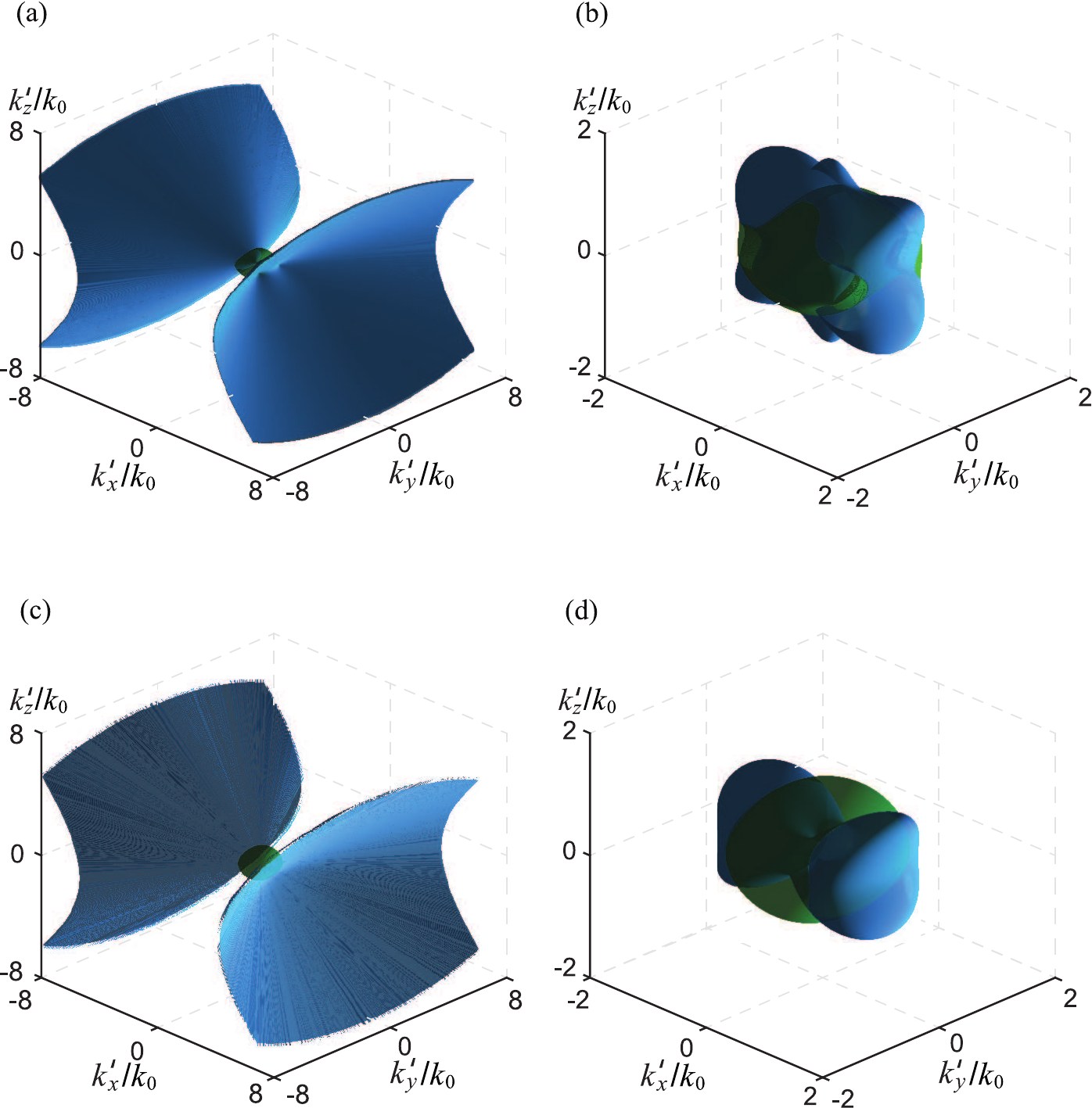}
\caption{Same as in Fig.~\ref{fig:fig5} but for conditions $0<\varepsilon_{zz}'<1$ and $-1<\mu_{xx}'<0$, and $\delta_m=0.2$. Parameters describing losses are: (a) $b = 1\times 10^{-4}$, $\nu = 1\times10^{-2}$~GHz; (b) $b = 5\times 10^{-2}$, $\nu = 1\times10^{-2}$~GHz; (c) $b = 1\times 10^{-4}$, $\nu = 5$~GHz; (d) $b = 5\times 10^{-2}$, $\nu = 5$~GHz.}
\label{fig:fig6}
\end{figure}

For the similar set of parameters, but when $\delta_m=0.2$ (Region III), an another combination $-1 < \mu_{xx}' < 0$ and $0 < \varepsilon_{zz}' < 1$ holds, and the superlattice behaves as an $\varepsilon$-near-zero medium. For such a medium the similar loss-induced topological transitions as well as forms of the isofrequency surfaces occur as for the previous case, except the revolution axis of the hyperboloid is oriented along the $x$-axis (Figs.~\ref{fig:fig6}(a) and \ref{fig:fig6}(c)).  

Since the given superlattice consists of magnetic and semiconductor subsystems, there is their competing influence when an external magnetic field is applied. It inevitably results in the topological behaviours which are distinctive from those typical for traditional hyperbolic metamaterials. Indeed, when $\varepsilon_{zz}'<0$ at a constant value of $b$, as $\nu$ arises, a non-trivial transition in the isofrequency surface of the extraordinary waves appears. The closed hyperbolic-like area gradually decreases in the direction of the $z$-axis, while it increases along both the $x$-axis and the $y$-axis. A complicate shape of the isofrequency surface of the extraordinary waves is conditioned by the loss-induced topological transitions from a Type I hyperboloid to a combination of two hyperboloids whose revolution axes are oriented orthogonally. Therefore, as  soon as losses in the semiconductor component increases, corresponding hyperbolic area becomes to be dominant resulting in a formation of the complicate bi-hyperbolic-like shape (Figs.~\ref{fig:fig5}(c) and \ref{fig:fig5}(d)). 

\begin{figure}[!t]
\centering
\includegraphics[width=0.95\linewidth]{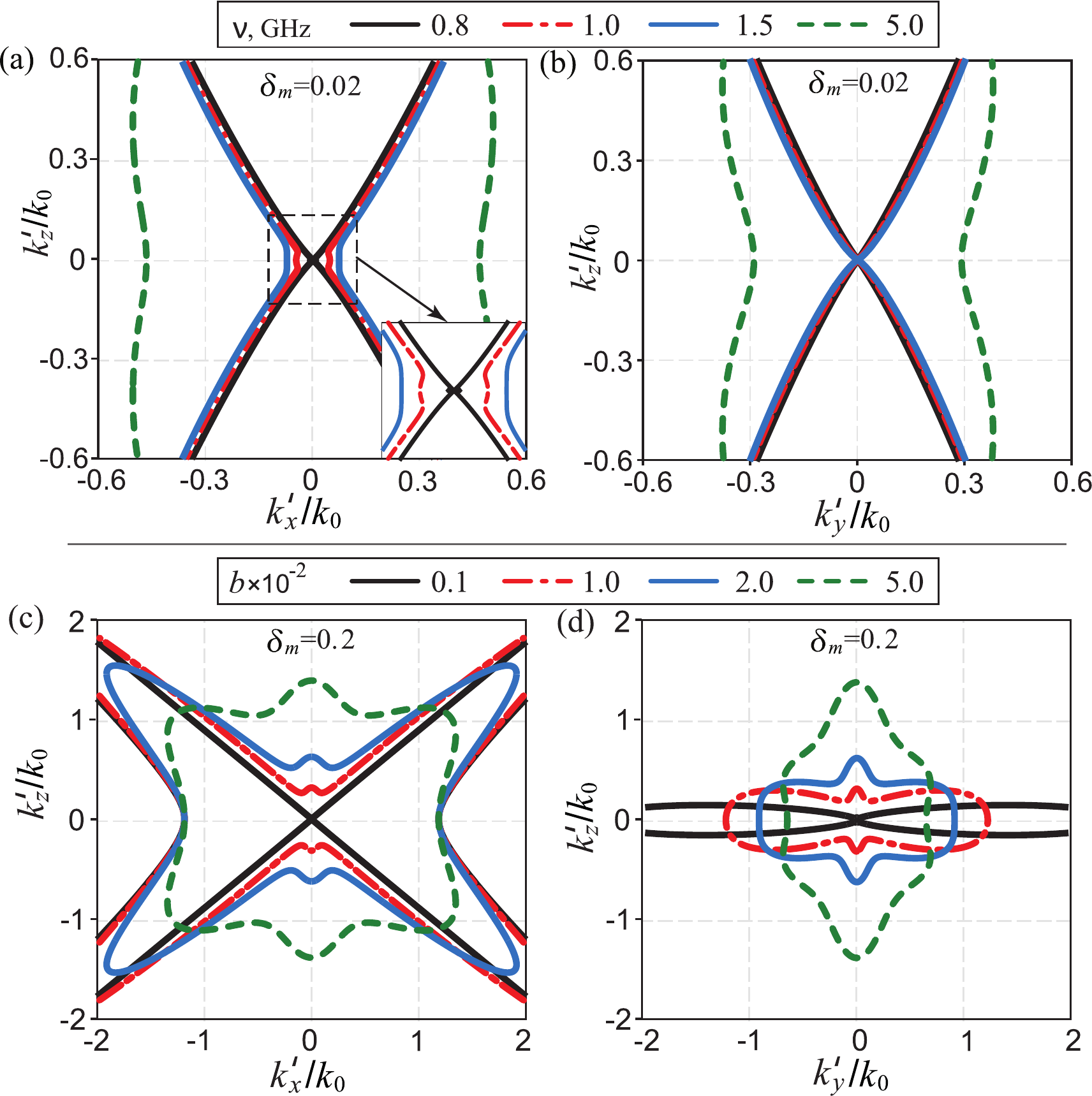}
\caption{The cross-section views of several stages of the loss-induced topological transitions of the isofrequency surface of the extraordinary waves; (a), (b) $\delta_m=0.02$ and $b = 2 \times 10^{-2}$; (c), (d) $\delta_m=0.2$ and $\nu = 1 \times 10^{-2}$~GHz.}
\label{fig:fig7}
\end{figure}

When $\mu_{xx}'<0$ at a constant value of $\nu$, as $b$ arises, the similar loss-induced transitions are observed in the superlattice where the magnetic component is dominant (Fig.~\ref{fig:fig6}(b)). In this case the closed isofrequency surface of the extraordinary wave undergoes compressing along both the $x$-axis and the $y$-axis and expanding in the $z$-axis direction. The variation of the closed area is simultaneously accompanied by shape changing of the isofrequency surface of the extraordinary waves (Figs.~\ref{fig:fig6}(b), \ref{fig:fig6}(d)). We have added Fig.~\ref{fig:fig7} with the cross-section views plotted at several discrete values of the parameters $b$ and $\nu$ to demonstrate these topological transitions in detail (an animation of these transitions occurring in the full range of parameters $b$ and $\nu$ one can find in the Supplemental Material\cite{Suppl_Mat})

\begin{figure}[!t]
\centering
\includegraphics[width=1.0\linewidth]{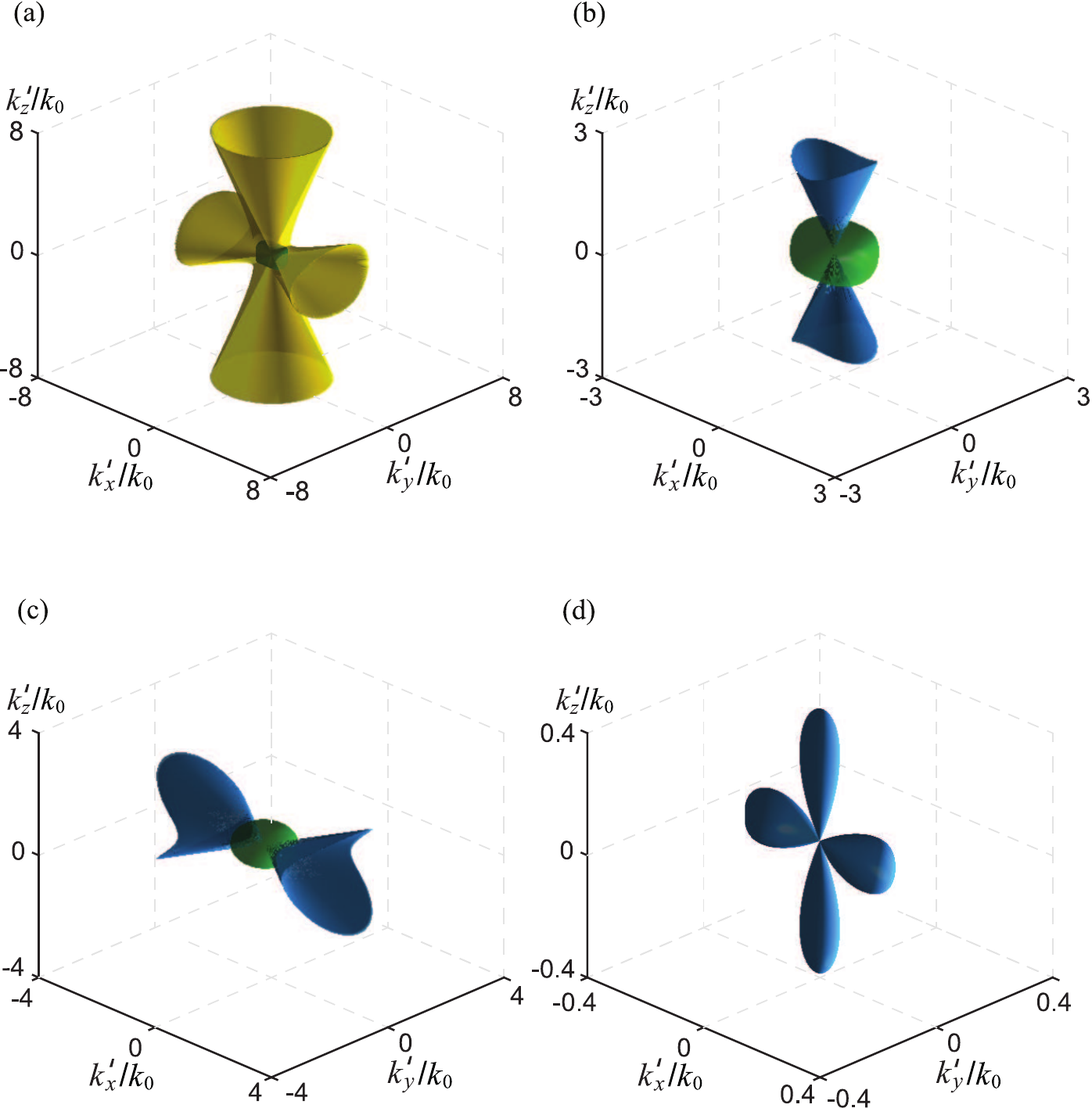}
\caption{Same as in Fig.~\ref{fig:fig5} but for conditions $-1<\varepsilon_{zz}'<0$ and $-1<\mu_{xx}'<0$, and $\delta_m=0.06$. Parameters describing losses are: (a) $b = 1\times 10^{-4}$, $\nu = 1\times10^{-2}$~GHz; (b) $b = 1\times 10^{-2}$, $\nu = 1\times10^{-2}$~GHz; (c) $b = 1\times 10^{-4}$, $\nu = 2$~GHz; (d) $b = 2\times 10^{-2}$, $\nu = 2$~GHz. Yellow surface corresponds to behaviours of extraordinary improper waves, whereas green and blue surfaces correspond to those of ordinary and extraordinary proper waves, respectively.}
\label{fig:fig8}
\end{figure}

The most distinctive is Region II, where a combination  $\mu_{xx}' < 0$ and $\varepsilon_{zz}' < 0$ of principal components of tensors $\hat\mu_{\textrm{eff}}$ and $\hat\varepsilon_{\textrm{eff}}$ holds. The superlattice behaves as an anisotropic double-negative medium for the extraordinary waves, while for the ordinary waves it is a double-positive one. For the extraordinary waves the isofrequency surface appears to be in a form of bi-hyperboloid,\cite{Tuz_OptLett_2017} which is significantly different from those obtained above for Regions I and III  (Fig.~\ref{fig:fig8}(a)). 

Although such an intriguing form of the isofrequency surface is found out, it is revealed that it corresponds to the nonphysical solutions of Eq.~(\ref{eq:disp_eq_k}) as soon as a small amount of losses is introduced into the system. That means the corresponding roots of Eq.~(\ref{eq:disp_eq_k}) describe the propagation of the improper waves. In Fig.~\ref{fig:fig8}(a) we plot the isofrequency surface for the extraordinary improper waves by yellow color.      

The most nontrivial effect is that the competition between the losses in magnetic and semiconductor subsystems leads to a change in the propagation conditions for the extraordinary waves. Thus, under certain level of either magnetic or semiconductor losses these waves become to be proper ones. In this case the isofrequency surface of the extraordinary waves acquires a loss-induced topological transition to a Type I hyperboloid whose revolution axis is oriented either along the $z$-axis (Fig.~\ref{fig:fig8}(b)) or the $x$-axis (Fig.~\ref{fig:fig8}(c)) depending on either magnetic ($\mu_{ij}'' \gg \varepsilon_{ij}''$ ) or semiconductor ($\varepsilon_{ij}'' \gg \mu_{ij}''$) subsystem is respectively dominant (an animation of these loss-induced transitions one can find in the Supplemental Material\cite{Suppl_Mat}). Thus, one may conclude that, in this region the dispersion conditions of the extraordinary waves can be significantly modified (up to orthogonal) by performing selective injection of losses into constitutive materials forming the superlattice. Moreover, if there is no competition between the losses in subsystems, then the loss-induced bi-hyperbolic-like topology arises (Fig.~\ref{fig:fig8}(d)).

\section{Conclusions}
\label{Concl}

We have studied the topological transitions of isofrequency surfaces existing in a magnetic-semiconductor superlattice which are influenced by an external static magnetic field. It is demonstrated that in the case of a lossless structure the topological transitions from a closed ellipsoid to open Type I and Type II hyperboloids as well as a bi-hyperboloid can be achieved. Such transitions of isofrequency surfaces occur at the critical points, where principal components of the effective permeability and/or permittivity tensors change their sign. 

The realization of the loss-induced topological transitions of isofrequency surfaces was studied in detail. We have demonstrated that several distinctive loss-induced topological transitions of the isofrequency surface of the extraordinary waves can be achieved by providing selective injection of losses into magnetic and semiconductor subsystems of the superlattice. For the first time it has been revealed that the loss-induced topological transition from a Type I hyperboloid to a bi-hyperboloid occurs when the real part of the principal component of permittivity and/or permeability is a near-zero quantity while its imaginary part is high enough.

%-----------------------------------------------------%
\section*{ACKNOWLEDGMENTS}
The authors acknowledge Jilin University's hospitality and financial support. 

\section*{Appendix. Constitutive parameters of ferrite and  semiconductor layers}
\label{AppA}
%--------------------------------------------------
\renewcommand{\theequation}{A.\arabic{equation}}
\setcounter{equation}{0}

The expressions for tensors components of the underlying constitutive parameters of magnetic $\hat \mu_m\to\hat g_m$ and semiconductor $\hat \varepsilon_s\to\hat g_s$ layers can be written in the form
\begin{equation}
\hat g_j=\left( {\begin{matrix}
   {g_1} & {\text{i}g_2} & {0} \cr
   {-\text{i}g_2} & {g_1} & {0} \cr
   {0} & {0} & {g_3} \cr
\end{matrix}
} \right). \label{eq:gfs}
\end{equation}

For magnetic layers \cite{Gurevich_book_1963, Collin_book_1992} the components of tensor $\hat g_m$ are $g_1=1+\chi' + \text{i}\chi''$, $\quad g_2=\Omega'+\text{i}\Omega''$, $ g_3=1$, and $\quad\chi'=\omega_0\omega_m[\omega^2_0-\omega^2(1-b^2)]D^{-1}$, $\chi''=\omega\omega_m b[\omega^2_0+\omega^2(1+b^2)]D^{-1}$, $\quad\Omega'=\omega\omega_m[\omega^2_0-\omega^2(1+b^2)]D^{-1}$, $\Omega''=2\omega^2\omega_0\omega_m bD^{-1}$, $\quad D=[\omega^2_0-\omega^2(1+b^2)]^2+4\omega^2_0\omega^2 b^2$, where $\omega_0$ is the Larmor frequency and $b$ is a dimensionless damping constant.

For semiconductor layers \cite{bass1997kinetic} the components of tensor $\hat g_s$ are $g_1=\varepsilon_l\left[ {1-\omega_p^2 (\omega+\text{i}\nu)[\omega((\omega+\text{i}\nu)^2-\omega_c^2)]^{-1}}\right]$, $\quad g_2=\varepsilon_l\omega_p^2\omega_c[\omega((\omega+\text{i}\nu)^2-\omega_c^2)]^{-1}$, $\quad g_3=\varepsilon_l\left[{1-\omega_p^2[\omega(\omega+\text{i}\nu)]^{-1}}\right]$, where $\varepsilon_l$ is the part of permittivity attributed to the lattice, $\omega_p$ is the plasma frequency, $\omega_c$ is the cyclotron frequency and $\nu$ is the electron collision frequency in plasma.

Relative permittivity $\varepsilon_m$ of the ferrite layers as well as relative permeability $\mu_s$ of the semiconductor layers are scalar quantities.

\bigskip

\bibliography{topological}

\end{document}